\documentclass[useAMS,usenatbib]{mn2e}
\usepackage{graphicx}

\def\deg{\ifmmode^\circ\else$^\circ$\fi}

\usepackage[latin1]{inputenc}
\usepackage[flushleft]{threeparttable}
\usepackage{booktabs}
\usepackage{titlesec}
\usepackage{lscape}
\usepackage{verbatim}
\usepackage{rotating}
\usepackage{multirow}
\usepackage{amsmath}	
\usepackage{amssymb}	
\usepackage{amsfonts}
\usepackage{lineno,hyperref}

\title[Star Cluster Detection using Parzen Density Estimation]{Star Cluster Detection and Characterization using Generalized Parzen Density Estimation}
\author[S. Nambiar et al.]{
Srirag Nambiar$^{1}$,
Soumyadeep Das$^{2}$,
Sarita Vig$^{1}$
and Gorthi R.K.S.S. Manyam$^{3}$\thanks{E-mail: rkg@iittp.ac.in}
\\
$^{1}$Indian Institute of Space science and Technology, Thiruvananthapuram 695547, India\\
$^{2}$Indian Institute of Technology (BHU), Varanasi 221005, India\\
$^{3}$Indian Institute of Technology, Tirupati 517506, India\\
}

\begin{document}

\date{}

\pagerange{\pageref{firstpage}--\pageref{lastpage}} \pubyear{}

\maketitle

\label{firstpage}

\begin{abstract}
Star cluster studies hold the key to understanding star formation, stellar evolution, and origin of galaxies. The detection and characterization of clusters depend on the underlying background density and the cluster richness. We examine the ability of the Parzen Density Estimation (a.k.a. Parzen Windows) method, which is a generalization of the well-known Star Count method, to detect clusters and measure their properties. We apply it on a range of simulated and real star fields, considering square and circular windows, with and without Gaussian kernel smoothing. Our method successfully identifies clusters and we suggest an optimal standard deviation of the Gaussian Parzen window for obtaining the best estimates of these parameters. Finally, we demonstrate that the Parzen Windows with Gaussian kernels are able to detect small clusters in regions of relatively high background density where the Star Count method fails.
\end{abstract}

\begin{keywords}
Open clusters and associations: general -- Methods: statistical -- Methods: numerical -- Techniques: miscellaneous
\end{keywords}

\section{Introduction}

A star cluster is defined as a conspicuous concentration of stars above the stellar background, that is localized in space.  This definition gives a prescription for foraging star clusters, by locating a system of objects protruding well above the background. The study of stellar clusters is important for a variety of reasons. Most stars are believed to have been formed in clusters and such clustered environment is likely to affect their pre-main sequence evolution \citep{2014prpl.conf..243K}.  In addition, these clusters can be used to test our understanding of mass segregation, stellar collisions and mergers, stellar evolution and core-collapse \citep{meylan2000001,2010RSPTA.368..755K,olczak2011001}. Furthermore, star clusters can help address  problems related to the origin of galaxies \citep{2008IAUS..246...13K,bica2016001,2016ApJ...828...75F}.

The members of a young star cluster are mostly coeval having formed from the same molecular cloud, hence observations of numerous stellar clusters in distinct stages of evolution can provide clues about the underlying mechanisms governing their formation and dynamics \citep{2011MNRAS.410L...6G}. Studies of young embedded clusters in molecular clouds located in the Solar neighbourhood \citep{ladalada2003001, lada2009001} have led to a coherent understanding of the initial mass function and star formation rate in our Galaxy \citep{2010ARA&A..48..339B}. Young embedded clusters are relatively difficult to isolate and characterise as they are obscured by the intervening dust from the parent molecular cloud along the line-of-sight \citep{2007prpl.conf..361A}. Near- and mid-infrared observations come to aid in such cases because of the reduced dust extinction, as compared to the optical regime \citep{schlafly2016001}. This has resulted in larger number of clusters being detected in infrared as compared to the optical \citep{ivanov2010001}.

The role of background for cluster detection cannot be understated, particularly in automated techniques that rely on statistical methods.  Clustering is noticeable so long as the stellar density is higher than that of the surroundings, which includes contribution from both, the foreground and background field stars. The background is usually estimated from regions of identical size in the neighbourhood of the cluster. More robust stellar background determination, using regions that are much larger in size (upto 4 times), in conjunction with alternate parameters has been explored by \citet{ivanov2017001}. The sensitivity of a catalog or survey as well as angular resolution are expected to play significant roles in the revelation of clusters. This is because deeper the survey, larger the contribution of stars from (i) the lower mass end of the Initial Mass Function (IMF), as well as (ii) the field stars (i.e. background). Further, high angular resolution would be instrumental in resolving out cluster members in regions of high stellar density of the cluster. These competing factors along with interstellar extinction determine whether the cluster overdensity is markedly visible above that of the neighbouring sky density. This is evident from the case of an embedded cluster in IRAS~20286+4105, a star-forming region located in the Cygnus complex. This sparse cluster was detected by \cite{kumar2006001} using the Two Micron All-Sky Survey (2MASS) data, and a revisit using deeper UKIRT Infrared Deep Sky Survey (UKIDSS; \citealt{lawrence2007001}) images by \cite{2017MNRAS.465.4753R} demonstrated that the cluster detection is suppressed due to the overwhelming background stellar density, as the sensitivity limit of UKIDSS is approached. Similarly, \citet{ryu2018001} reported 923 new clusters using Wide-field Infrared Survey Explorer (WISE) data \citep{wright2010001}, which uses longer wavelength as compared to UKIDSS, in an area where only 339 clusters were known previously.

The statistical methods used for stellar cluster detection and characterisation employ density estimation techniques. The probability density estimation is one such method and in its most basic form follows the approach of Parzen Windows. In this method, the area under observation is divided into overlapping windows of equal area and the stellar probability densities in the windows compared. A special case of this approach that uses square windows is commonly known as the Star Count method. Typically, a stellar Probability Density Function (PDF) contour map of the region under consideration is constructed and the presence of a cluster is established through contour levels higher than the surrounding background density. There are two basic categories of approaches for density estimation in statistical pattern recognition, viz., \emph{parametric} and \emph{non-parametric} density estimation techniques. Parametric methods are useful if the mathematical form of the cluster PDF is known beforehand. But since nature denies us this luxury, we rely on non-parametric methods. 

The most traditional, simplest and widely used  method used for stellar density estimation has been the Star Count method.  Numerous studies, for instance: \citet{lucas2008001}, \citet{kirsanova2008001}, \citet{karampelas2009001}, \citet{shmeja2014001} including recent ones like \citet{hony2015001}, \citet{belokurov2016001}, \citet{constantino2016001}, \citet{fritz2016001} and \citet{gallego2017001} have used the Star Count approach to investigate clustering and other properties of the clusters. Alternate methods used for stellar cluster identification include visually searching for cluster density enhancements \citep{borissova2014001}, the Poisson model based algorithm \citep{mercer2005001}, k-Nearest Neighbour method, Voronoi tessellation, and minimum spanning tree separation \citep{schmeja2011001}. 

Our focus in the present work is to use a simple but effective algorithm to detect young embedded clusters, particularly in regions of dense stellar background. We develop a systematic framework to detect such stellar clusters using the well developed theory of Parzen Windows that is quite general in the sense that it is easily applicable to any area where density estimation is essential.  

The paper is organized in the following manner. In Sections 2 and 3, we introduce the Parzen Window approach and provide the details of simulated and real clusters, respectively. In Section 4, we describe the proposed methodology. The results of simulated and real clusters are elucidated in Sections 5 and 6, respectively. Finally, we present the conclusions in Section 7. 

\section{Parzen Windows for Star Cluster Density Estimation}

In statistical pattern recognition, the methodology of Parzen Windows is the most straight-forward technique for density estimation. In this method, the area under investigation is divided into small overlapping bins or windows. The identical windows could be of any shape. We have used square and circular windows in the current work. The total region under consideration is assumed to have $N$ points, where each point represents a star. The PDF of the $i^{th}$ window, $P_{i}$, is a function of $k_i$, the number of points falling in that window. If the PDF of certain windows is above a given threshold associated with the background, then a cluster is said to be detected. We list below the different forms of windows considered in the present work.
\begin{itemize}
 \item[i.] \textbf{Square Parzen Window or Star Count:} In this method, the shape of the windows is square and equal weightage is given to all points lying within the window. The total area is covered with squares of side length $d$, with an overlap of $d/2$, i.e. half the side length of two adjacent square windows. This window overlap ensures that the method is in accordance with the Nyquist sampling criterion. The probability density, $P_{i}$, in this case, is given by 
 \begin{equation}
 P_{i} =\frac{k_i}{Nd^2 }
 \label{eq1}
 \end{equation}
Throughout the paper, we shall use the names Star Count and square Parzen Window approach interchangeably, as they are identical.
\vspace*{0.3cm}
\item[ii.] \textbf{Circular Parzen Window:} This is similar to Case (i) where equal weightage is given to all the points in a given window. However, the shape of the window is circular unlike the previous case. The total area is covered using circles of radius \emph{r} and their centres separated by \emph{r} (half the side length of square windows assuming $r=d/2$). Again, this results in an overlap of \emph{r} between adjacent windows. The probability density in the $i\textsuperscript{th}$ circular window is written as 
 \begin{equation}
 P_{i} = \frac{k_i}{N\pi r^2 }
  \label{eq2}
 \end{equation} 
\vspace*{0.3cm}
\item[iii.] \textbf{Circular Gaussian Window:} In this method, the points within a given window are assigned certain weights based on their location from the center of the window. This is unlike Cases (i) and (ii), where equal weightage is given to all points irrespective of their locations, leading to a disadvantage due to flattening of the PDF. Assigning weights is achieved by using Smooth Kernel functions. The most common Kernel function is the Gaussian Kernel and a smoothing parameter called $\sigma$ is associated with the windows that decides the weightage attributed to a point according to its distance from the center of the window $(\overline{x},\overline{y})$. In other words, the normalised Gaussian is centred on the central point of the circular window. The shape of the window is circular similar to Case (ii), and the radius of the circle is $3\sigma$ (based on the property of Gaussian distribution). The overlap between windows is again one radius.

The selection of $\sigma$ for real and simulated clusters is discussed in Section 4.1. The contribution of the $j^{th}$ point, located at $(x_j,y_j)$, to the PDF estimate of the window depends on the separation between $(x_j,y_j)$ and $(\overline{x},\overline{y})$. The probability density function $P_i$, of this kernel is given by 
 \begin{equation}
 P_{i} =\sum_{j=1}^{k_i}\frac{1}{2\pi\sigma^2 } e^\frac{-(x_{j}-\overline{x})^2-(y_{j}-\overline{y})^2}{2\sigma^2}
  \label{eq3}
 \end{equation} 
\vspace*{0.3cm}
\item[iv.] \textbf{Square Gaussian Window:} This case is similar to Case (iii) except for the fact that the shape of the window is square. The side length of each square is 6$\sigma$ with an overlap of 3$\sigma$ between the adjacent windows. The probability density for the $i\textsuperscript{th}$  square window is given as \\
 \begin{equation}
 P_{i} =\sum_{j=1}^{k_i}\frac{1}{2\pi\sigma^2 } e^\frac{-(x_{j}-\overline{x})^2-(y{j}-\overline{y})^2}{2\sigma^2}
  \label{eq4}
 \end{equation} 
\vspace*{0.3cm}
\item[v.] \textbf{k-Nearest Neighbour approach:} This is also a window based density estimation technique. However, unlike the previous cases, where the size of the window is fixed, in this case the window size varies inversely with local density. This is achieved by establishing a pre-determined number of neighbours `$k$' and varying the size of the square window $d_i$ to include them. Here equal weightage is given to all points lying within the window. The probability density, $P_{i}$, in this case is given by
 \begin{equation}
 P_{i} =\frac{k}{Nd_i^2 }
  \label{eq5}
 \end{equation} 

Please refer to \cite{schmeja2011001} for further details on this approach and its application to stellar density estimation.  
\end{itemize}

\section{Cluster Sample}
In order to detect stellar clusters and characterise them using the PDF algorithms described earlier, we follow a two step approach. In the first step, we simulate clusters of diverse morphologies in order to compare the performance of the algorithms and determine if any specific algorithm outperforms others in all cases. The morphologies of simulated clusters are approximately driven by the appearance of observed star clusters. In the second step, we proceed to apply these algorithms to few cases of real stellar clusters as well as regions where clustering is expected based on other considerations.  


\begin{figure*}
    \centering
    \includegraphics[width=\textwidth]{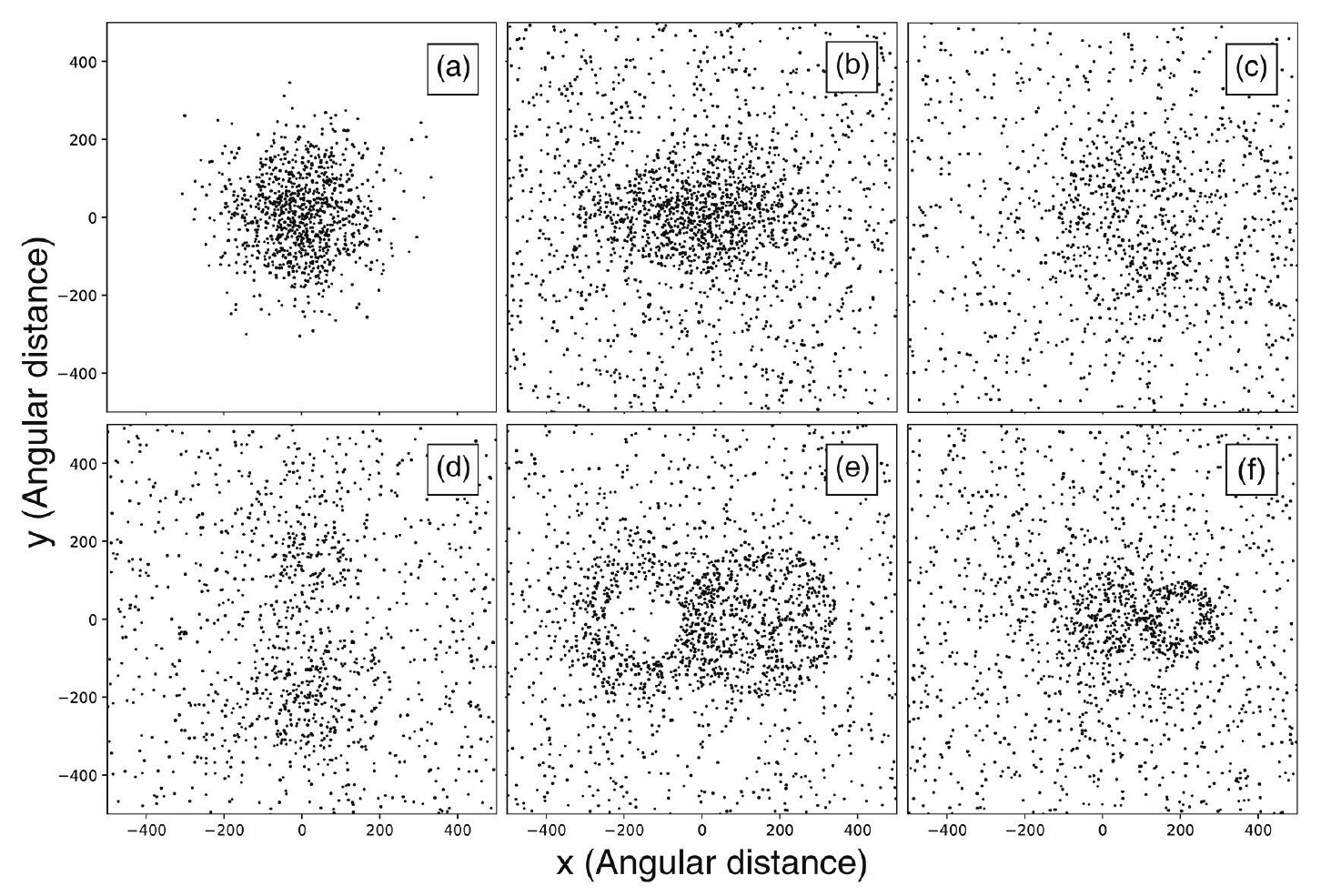}
    \caption{Simulated Clusters: (a) S01, (b) S02, (c) S03, (d) S04, (e) S05, (f) S06. The details of these clusters are given in Table~\ref{Table:Sim_Clus}. }
    \label{Sim_Clus}
\end{figure*} 


\subsection{Simulated Clusters}

To test the behaviour of the algorithms under known conditions, different sets of model clusters are generated which reflect the wide variety of clusters found in nature. Broadly, real star clusters are categorised into two types based on their structure: (i) Centrally condensed type star clusters, and (ii) Hierarchical-type clusters. The structure of a cluster is related to the underlying star formation mechanisms at work \citep{ladalada2003001, guter2005001,domin2017001}. Centrally-condensed type of clusters usually have a single prominent peak and show highly concentrated surface density distributions with relatively smooth radial profiles, whereas hierarchical-type clusters display evidence of clustering over a large area and contain multiple peaks. In some cases, massive centrally concentrated clusters  often appear as rings or ``doughnuts". Such ring or shell type clusters could be the result of high extinction or confusion of sources in the central region \citep{schmeja2011001}. Alternately, the ring structures could be the outcome of the underlying gas density distribution with a central cavity \citep{mathieu2008001,kumar2003001}. 


\begin{table*}
  	\centering
  	\caption{Details of Simulated Clusters.}
  	\medskip
  	
    \begin{tabular}{rclcc}
    \toprule
    S.No. & Cluster Name & Type  & Stars in Cluster & Stars in Background \\
    \midrule
    1 & S01 & Single Gaussian Cluster without Background noise & 1000  & 0 \\
    2 & S02 & Gaussian cluster with Uniform noise & 400   & 100 \\
    3 & S03 & Gaussian and Circular clusters with Non-uniform noise & 540   & 700 \\
    4 & S04  & Two Gaussian Clusters with Non-uniform noise & 500   & 700 \\
    5 & S05 & Circular and Gaussian Doughnut Clusters with Uniform noise & 1016  & 800 \\
    6 & S06 & Gaussian and Circular Doughnut Clusters with Non-uniform noise & 607   & 1000 \\
    \bottomrule
    \label{Table:Sim_Clus}
    \end{tabular}
\end{table*}


Four basic simulated clusters, viz., \emph{Gaussian, Circular, Gaussian Doughnut} and \emph{Circular Doughnut} are created by employing uniform random number and multi-variate random number distributions in Matlab\textsuperscript{\textregistered}. Combining them together and adding noise using uniform random number distributions in Matlab\textsuperscript{\textregistered}, six simulated clusters are generated. The noise, which is a reflection of the background stellar density, may or may not be uniform. While we have simulated uniform background in few cases, we have also constructed non-uniform background distributions using spatially segregated uniform star distributions.  A brief description of the simulated clusters is given in Table~\ref{Table:Sim_Clus} while their graphical visualisation is shown in Figure ~\ref{Sim_Clus}.


\begin{figure*}
	\centering
	\includegraphics[width=\textwidth]{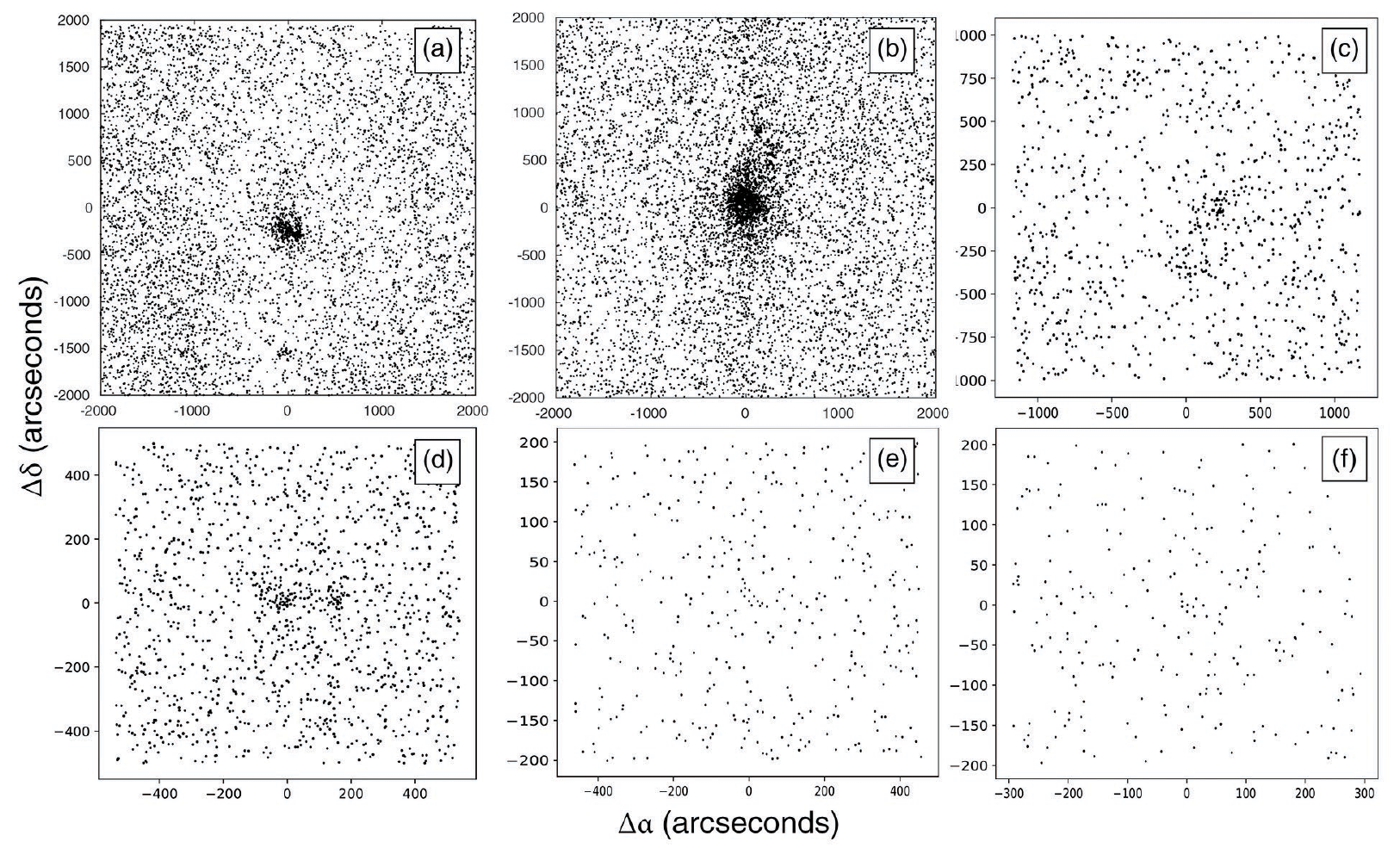}
	\caption{Real Clusters: (a) NGC 2024 (b) Trapezium (c) NGC 1333 (d) IRAS 06061+2151 (e) IRAS 01420+6401 (f) IRAS 04579+4703. The cluster centers, corresponding to the origin of the plots, are listed in Table~\ref{Table:Real_Clus_results}.}
    \label{Real_Clus}
\end{figure*} 


\subsection{Real Clusters}
We have tested our algorithms on real clusters, and for our sample we have selected a total of six regions that either have or are expected to host clusters. Our selection method is broadly based on the evolutionary state of the molecular cloud, that has direct implications on the detection of the cluster as well as its membership status. Furthermore, we have attempted to include clusters with distinct morphological features. The less evolved and younger embedded clusters (such as NGC 2024, NGC 1333) are found in massive dense molecular cores, while the more evolved clusters (e.g., the Trapezium) are located within HII regions and reflection nebulae at the edge of molecular clouds. Based on morphology, the hierarchical-type clusters include the deeply embedded double cluster NGC 1333 \citep{Lada1996} and IRAS 06061+2151 \citep{molinary1996001, kumar2006001}.  M42 (Trapezium) and NGC 2024 \citep{ladalada2003001} belong to the category of centrally condensed type of cluster. We have also included two regions, IRAS~01420+6401 and IRAS~04579+4703, where embedded clusters are expected based on other tracers of star formation such as IRAS colours, ammonia emission and water masers \citep{molinary1996001}. These two star-forming regions were among those, where embedded clusters were not detected by \cite{kumar2006001} using the Star Count method. These authors considered a set of 217 targets listed by \cite{molinary1996001} and \cite{sri2002001} and detected 54 clusters out of them. The low detection rate of $\sim25$\% is attributed to high extinction as well as background stellar overdensity associated with the Galactic plane.

Our sample comprises an assorted mix of young embedded clusters with varying memberships. To carry out the analysis, we use stellar sources detected in the K-band (2.2$\mu$m) as it suffers relatively low extinction. For each  real cluster, the positional information (Right Ascension and Declination) of the sources in a given box was obtained from the Two Micron All-Sky Survey (2MASS) catalog \citep{skrutskie2006001}. The regions of sky that have been analysed for cluster detection, are shown in Fig.~\ref{Real_Clus} with the cluster centres listed in Table~\ref{Table:Real_Clus_results}.

\section{Proposed Methodology}

The detection of  clustering by the method of Parzen windows is  a three-step process, as explained below: 
\begin{itemize}
 \item [I.] \textbf{Defining the Window size:} The window size is selected such that the number of stars per window is neither too large to miss an overdensity, nor too small to get a false detection. We use a single common parameter $dx$ to define the window size for the various cases considered. For square Parzen windows, the window size is $dx\times dx$ and they are separated by $dx/2$ units \citep{Lada1996, kumar2006001, schmeja2011001}. In case of  circular Parzen windows, circular regions or windows are of diameter $dx$, separated by $dx/2$ units. In the case of circular Gaussian windows, the smoothing parameter $\sigma$  of the window is taken to be $dx/6$, with a circular area of radius $3\sigma$ serving as the Parzen window. Similarly for square Gaussian windows, $\sigma$ is taken as $dx/6$ while squares of side 6$\sigma$ serve as Parzen windows. A more detailed description regarding the selection of optimal window size for a given field of sky is discussed in section 4.2.


\begin{figure*}
    \centering
    \includegraphics[width=\textwidth]{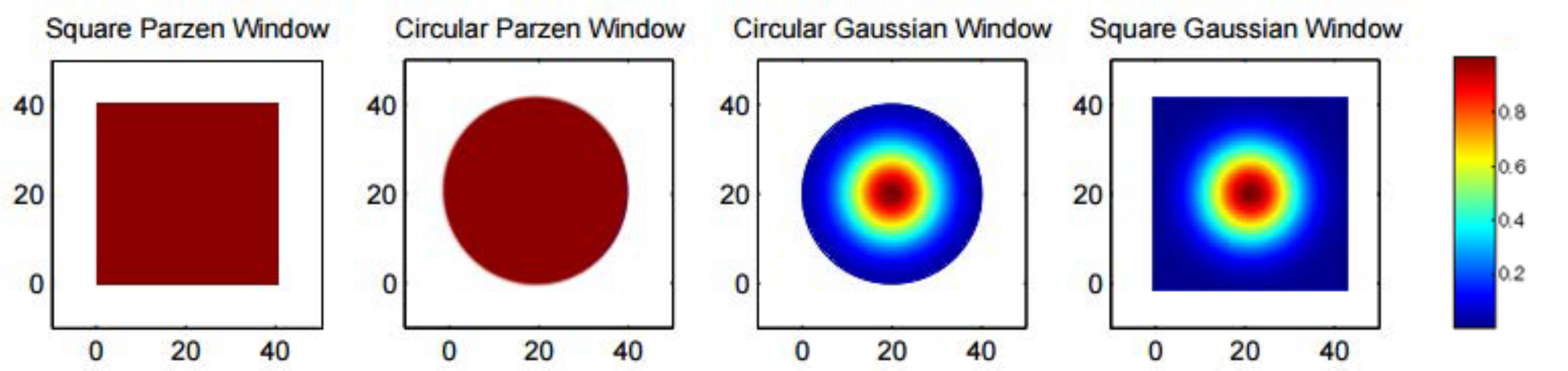}
    \caption{Probability Distribution of Parzen windows with $dx = 40$ (for Gaussian window, $\sigma  = 40/6$). The colour represents the probability density function, defined as the fraction of stars (lying around each point) to the total number of stars. See text for additional details.}
    \label{Windows_PDF}
\end{figure*} 



\begin{figure}
    \centering
    \includegraphics[scale=0.24]{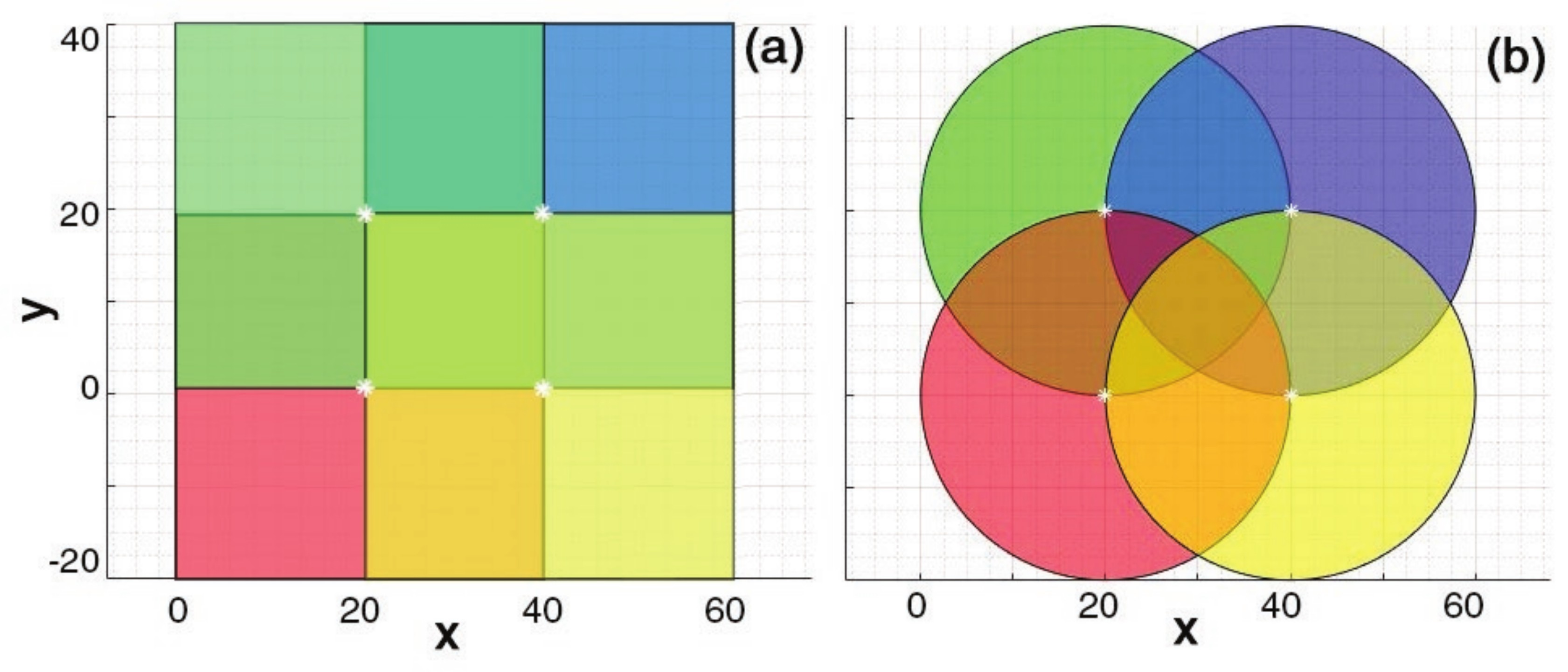}
    \caption{Overlapping arrangement of (a) Square, and (b) Circular Parzen windows. The white markers denote window centers.}  
    \label{Windows}
\end{figure} 


The sizes of real star clusters that we investigate are in the range $100''-1000''$ with smaller scale structures within. We use square Parzen windows with a window size of $80'' \times  80''$ and a step size of $40''$ to map the morphological details, for real clusters. Circular Parzen windows with radius $40''$ is considered. Circular and square Gaussian windows are considered such that $3\sigma = 40''$. The arrangement of windows is shown in Figure~\ref{Windows} for $dx=40$. For simulated clusters, we proceed with small clusters within a region of size $1000''$. It is to be noted that although we have assigned an incremental unit of simulations to 1 arcsec for simplicity, the simulations are general enough to be applicable to larger or smaller angular scales. 
\vspace*{0.3cm}
\item[II.] \textbf{Computing the PDF:} The PDF is a measure of how densely stars are distributed in a given region. We use a normalized Matlab\textsuperscript{\textregistered} histogram to generate a two-dimensional field, that serves as a standard with which to compare our results. The coordinates of the Parzen window centers are stored in a matrix $C$. The PDF is calculated for each window and stored in a matrix $P$.  In case of simulated clusters, the standard deviation ($\sigma_{est}$) of the detected cluster is estimated from the covariance matrix, which is obtained using the following expression: 
\begin{equation}
[Covariance]_{2 \times 2}=\frac{1}{N}  \sum_j(C_j - \mu)(C_j - \mu)^tP_j
 \label{eq6}
\end{equation}

Here, $N$ = Total number of stars, $P_j$ is PDF of $j^{th}$ window, and $\mu$ is a $1\times 2$ matrix storing the mean of the two-dimensional coordinates of the stars that we label X and Y for simplicity.

In a non-symmetric case, the diagonal elements of the covariance matrix represent the standard deviation in X($\sigma_x$) and Y($\sigma_y$). The absolute element-wise difference between the normalized Matlab\textsuperscript{\textregistered} histogram and the Parzen window based normalized PDF matrix gives us the error in our approximation.
\begin{equation}
Error = \sqrt{\sum_i[P(i)-P_{hist}(i)]^2}
 \label{eq7}
\end{equation}

A surface plot of the element-wise error for simulated cluster S02 is shown in Fig. \ref{hist_error}, as an example. If the standard deviation of the true stellar distribution is represented by $\sigma_{tr}$, the ratio of $\sigma_{est}$ to $\sigma_{tr}$  gives an additional measure of the proximity between the simulated standard deviation and that extracted by the algorithm. These two quantities are plotted as a function of (a) number of Parzen Windows, and (b) number of stars in a cluster, to judge the behavior of the algorithm. A point worth noting here is that $\sigma_{est} / \sigma_{tr}$ ratio is a useful measure only if the simulated cluster is a single Gaussian without background noise, S01 (discussed later in Sect. 5.1.1). This is because in the presence of noise, $\sigma_{tr}$ is a user defined value attributed to Gaussian alone whereas $\sigma_{est}$ comes from the combination of Gaussian and noise distribution. 

\begin{figure*}
    \centering
    \includegraphics[scale=0.4]{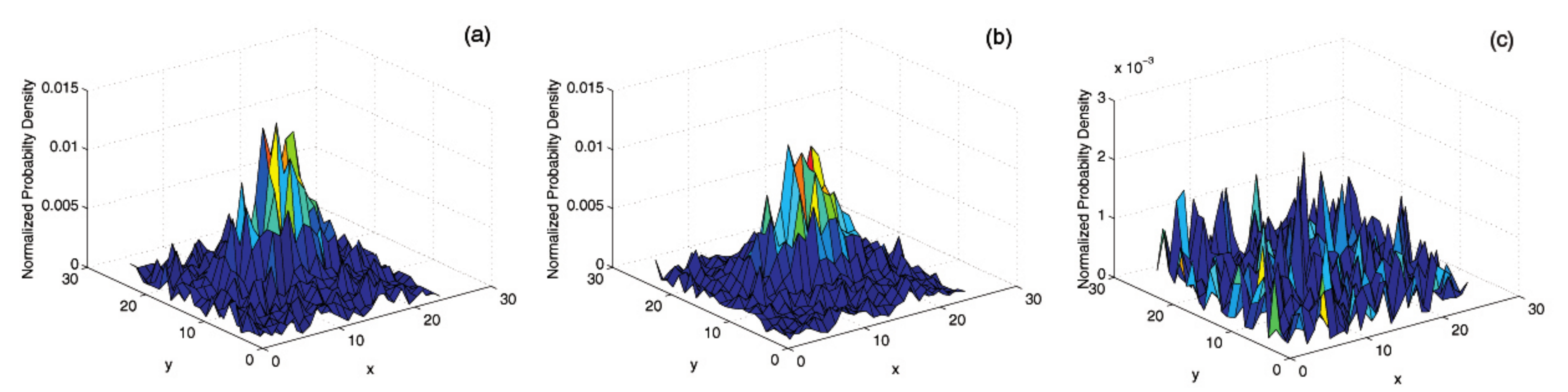}
    \caption{Surface plots of (A) Matlab\textsuperscript{\textregistered} Normalized Histogram (B) Normalized PDF (C) Error (Absolute element-wise difference b/w PDF and hist) for simulated cluster S02}
    \label{hist_error}
\end{figure*} 

\vspace*{0.3cm}
\item[III.] \textbf{Retrieving the Cluster:} 
The critical part of cluster identification is fixing a threshold above the background,  that can be used to qualify an overdensity as a potential cluster. Previous studies such as those by \citet{schmeja2011001}, \citet{kumar2006001}, \citet{mallick2015001} and \citet{baug2015001}, have considered regions adjacent to the cluster to define the background level. As mentioned earlier, this is relatively arbitrary. We present a more sophisticated and systematic approach for deriving this cut-off level. This is achieved by the following steps:
\begin{itemize}

\item Estimate an average foreground + background level.\\
The mode, i.e. the most frequently occurring value, in the normalized histogram decides an approximate foreground + background level in the field containing the cluster. This represents the average star count in the background, or the noise. The fluctuations in noise are defined with respect to this average level.
\vspace*{0.3cm}
\item Estimate the fluctuations in the background. \\
In order to estimate the background fluctuations about the mode, denoted by the parameter $\sigma_{bg}$, we consider the local maxima in the PDF. A local maximum is defined as the PDF corresponding to a certain window that lies above the PDF values of its immediate neighbouring windows, i.e. it forms a peak in its vicinity. An additional requirement is that the local maximum should lie above the background mode. Thus, an explicit definition of a PDF of window $(m,n)$, $P_{mn}$, to be a $i$\textsuperscript{th} local maximum, $L_i$, entails the satisfaction of the following conditions: 


\begin{figure}
    \centering
    \includegraphics[width=0.445\textwidth]{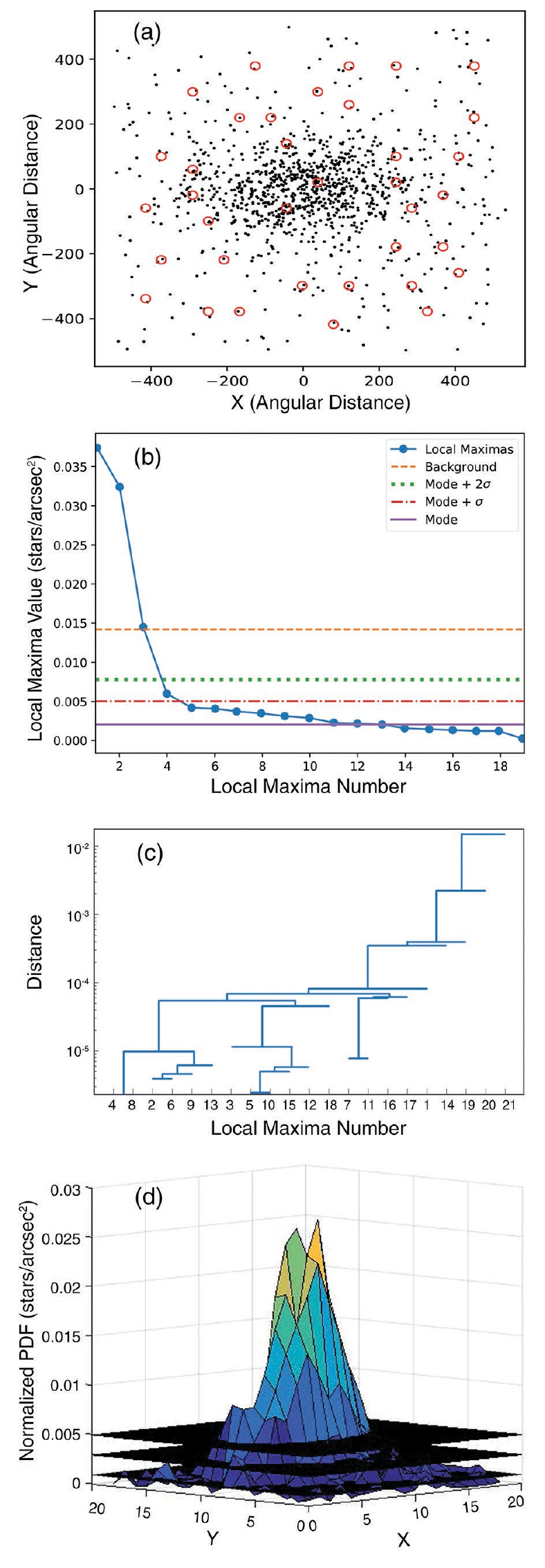}    
    \caption{Results of simulations for cluster S02: (a) Positions of Local maxima, (b) The Local maxima plot, (c) The Dendrogram plot, and (d) PDF displaying levels corresponding to mode,  mode+$\sigma$ \& mode+$2\sigma$ from bottom.}
    \label{locmax_mode_sigma}
\end{figure} 


\indent (i) $P_{mn} \ge P_{hk}$,\hspace{0.1cm} $h \hspace{0.08cm} \epsilon$ ($m-2$,$m+2$) and $k \hspace{0.08cm} \epsilon$ ($n-2$,$n+2$). \\
In a neighbourhood consisting of its second nearest neighbours (i.e. $5 \times 5$ neighbourhood), the Parzen window associated with $(m,n)$ has the maximum PDF value. \\
\indent (ii) $L_i \ge Mode$,\hspace{0.3cm}$\forall i$. \\ 
{This auxiliary condition restricts the number of local maxima to increase computational efficiency, as it is  futile to consider the large number of local maxima lying below the background mode.}

The local maxima indicating local stellar overdensities are illustrated as red circles in Figure~\ref{locmax_mode_sigma}(a). In order to estimate $\sigma_{bg}$, it is essential to exclude the local maxima associated with the cluster. This necessitates a local maximum limit ($\lambda$) that  can be used to estimate $\sigma_{bg}$. To determine $\lambda$, all the local maxima are plotted in a descending order, referred to as the local maxima curve. This is shown in Figure~\ref{locmax_mode_sigma}(b). Since the PDF values associated with a cluster are expected to be reasonably higher than the local maxima associated with the background, the local maxima curve shows an abrupt decrease at a certain local maximum value. The position of this abrupt change can be inferred by (i) a  visual selection, or (ii) an automated technique that relies on single linkage hierarchical clustering of $\Delta L = L_i - L_{i-1}$. Single linkage clustering is a hierarchical method, wherein two groups merge if their properties are very similar.

We follow the second approach and use the dendrogram method in order to investigate the grouping of local maxima to determine $\lambda$. A dendrogram is a tree or branch diagram used to illustrate the arrangement of local maxima produced in hierarchical clustering, where the branches represent different groups of local maxima. The linkage dendrogram plot that is used to extract the abrupt change is shown in Fig.~\ref{locmax_mode_sigma}(c). The Y-axis of the dendrogram features the property used for similarity measurement, in this case $\Delta L_i$, whereas a serial number given to each local maximum is plotted on the X-axis. The dendrogram plot can be used to estimate $\lambda$.
$\sigma_{bg}$ is then defined as the standard deviation of the local maxima lower than $\lambda$, with respect to the average background level or mode, in our case. 

\begin{equation}
\sigma_{bg} = \sqrt{\sum_i{\frac{(L_i - mode)^2 }{N}}},\hspace{0.5cm} L_i \leq \lambda
 \label{eq8}
\end{equation} 
\vspace*{0.3cm}
\item Identification of the Cluster.\\
Any collection of windows is identified as a cluster if their PDF values are greater than mode+$2\sigma_{bg}$. We visualise the cluster using contours of values mode+$2\sigma_{bg}$ and higher. Figure~\ref{locmax_mode_sigma}(d) illustrates the example of PDF values through a surface plot, with levels indicating the mode, mode+$\sigma_{bg}$ and mode+$2\sigma_{bg}$. 

\end{itemize}

\end{itemize}

\subsection{Cluster Properties}
Once the cluster is identified, we can determine various attributes of the stellar cluster. In case of clusters deeply embedded in molecular clouds, the spatial distribution of stars closely resembles the molecular cloud morphology at earlier stages and is less likely to be a result of dynamical evolution. If information about the mass of stars is available, then the problem of mass segregation in a cluster can be investigated \citep{parker2015001}. For cases where radial velocity measurements of stars are available, clustering in the third dimension (velocity) can also be analysed enabling the separation of superimposed clusters from the background accurately \citep{alfaro2016001}. Clusters with extended corona have also been illustrated using radial density profiles by \citet{seleznev2016001}. \citet{kumar2006001} have derived the mass, morphological type, photometry and extinction for the embedded clusters detected using the Star Count method. But estimating all these properties of clusters is involved and requires additional information which is out of scope of the current work. Here, we derive few basic attributes for the following reasons: (i) to compare the performance of different types of Parzen Windows, and (ii) to compare the properties of real clusters with those obtained from literature. These properties are discussed below in brief.
 

\begin{figure}
    \centering
    \includegraphics[width=.45\textwidth]{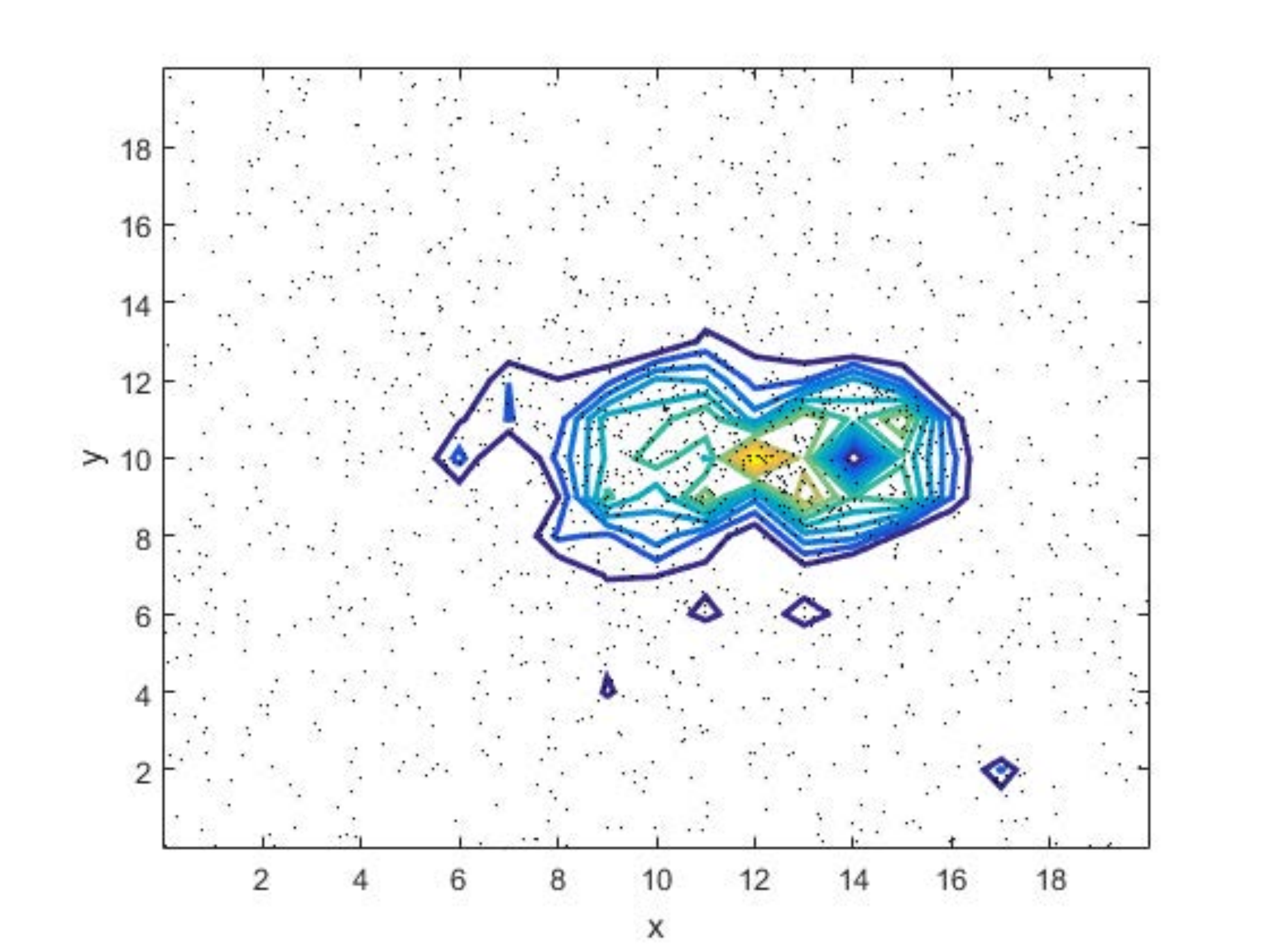}
    \includegraphics[width=.35\textwidth]{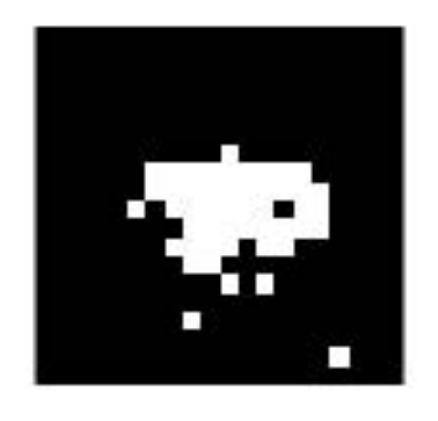}
    \caption{The top panel represents a contour plot for the simulated cluster S06, where the outermost contour corresponds to mode+$2\sigma$. The bottom panel displays the corresponding PDF mask.}
    \label{contour_map}
\end{figure} 

\begin{itemize}


\item Effective radius of the Cluster \\
The cluster is assumed to lie in a region defined by the contour level mode+$2\sigma_{bg}$.  This can be translated to an effective radius by equating the area to a circle. This parameter is useful for comparison with sizes of real clusters from literature.

 \item Cluster Membership\\
A logical approach for approximating the number of stars belonging to the cluster, i.e. cluster membership, is to consider the number of stars within the cluster area. To compute this, we estimate the total number of stars within a region enclosed by the contour level, mode+$2\sigma_{bg}$ level. However, such an estimate also includes stars associated with the background. Eliminating the contribution of stellar background from the total number of stars facilitates a fair idea of the cluster membership.

\begin{table*}
  \centering
  \caption{True and derived estimates of cluster membership from simulations}.
  \medskip
    \begin{tabular}{|c|c|c|rrrrr|}
    \toprule
    \multirow{2}{*}{S.No.} & \multirow{2}{*}{Cluster Name} & \multirow{2}{*}{No of Stars}  &  \multicolumn{4}{c|}{Cluster membership ($\%$ error)} \\
          &       &       & \multicolumn{1}{|c}{Square} & \multicolumn{1}{c}{Circle} & \multicolumn{1}{c}{Sq. Gauss.} & \multicolumn{1}{c|}{Cir. Gauss.} & \multicolumn{1}{c|}{kNN}\\
    \midrule
    1     & S02 & 400   & \multicolumn{1}{|c}{339 (15.25)} & \multicolumn{1}{c}{353 (11.75)} & \multicolumn{1}{c}{361 (9.75)} & \multicolumn{1}{c|}{361 (9.75)} & \multicolumn{1}{c|}{268 (33.00)}\\
    2     & S03 & 540   & \multicolumn{1}{|c}{530 (1.85)} & \multicolumn{1}{c}{516 (4.44)} & \multicolumn{1}{c}{545 (0.92)} & \multicolumn{1}{c|}{538 (0.37)} & \multicolumn{1}{c|}{424 (21.48)}\\
    3     & S04 & 500   & \multicolumn{1}{|c}{439 (12.2)} & \multicolumn{1}{c}{521 (4.20)} & \multicolumn{1}{c}{504 (0.80)} & \multicolumn{1}{c|}{499 (0.20)} & \multicolumn{1}{c|}{349 (30.20)}\\
    4     & S05 & 1016  & \multicolumn{1}{|c}{1196 (17.71)} & \multicolumn{1}{c}{1186 (16.73)} & \multicolumn{1}{c}{1065 (4.82)} & \multicolumn{1}{c|}{1065 (4.82)} & \multicolumn{1}{c|}{760 (25.20)}\\
    5     & S06 & 607   & \multicolumn{1}{|c}{695 (14.49)} & \multicolumn{1}{c}{654 (7.74)} & \multicolumn{1}{c}{612 (0.82)} & \multicolumn{1}{c|}{612 (0.82)} & \multicolumn{1}{c|}{551 (9.23)}\\
    \bottomrule
    \end{tabular}
    \label{Table:Sim_Clus_error}
\end{table*}

 \item Cluster Center\\ 
The center of a cluster can be inferred by taking the mean coordinates of members lying within the cluster area. If multiple clusters are detected then a center for each is obtained. For cluster configurations that possess a hierarchical structure, it is difficult to ascertain a unique cluster centre. In these cases too, we follow the same approach, as it gives a primary location for cluster identification.

\end{itemize}

\subsection{Pointers to select Window Size}
The selection of window size for cluster detection largely depends on the stellar density. In order to envision a relationship between cluster density and the window size, Gaussian clusters with varying densities were analysed. These simulations included  twenty single Gaussian cluster without any noise. The density was altered in two ways: (i) by varying the cluster membership and keeping the size constant, and (ii) by changing the cluster radius while keeping the membership fixed. In the first case, the number of stars was varied between 500 and 10000 within a cluster having a radius of $100''$. In the second case, the cluster radius was varied between $50''$ and $800''$ for a cluster membership of 1000. The Optimal Window Size (OWS) for each of these clusters was computed. The OWS is defined as the window size at which the error in cluster estimation as compared to the MATLAB\textsuperscript{\textregistered} histogram is minimum.

\begin{figure}
    \centering
    \includegraphics[width=0.45\textwidth]{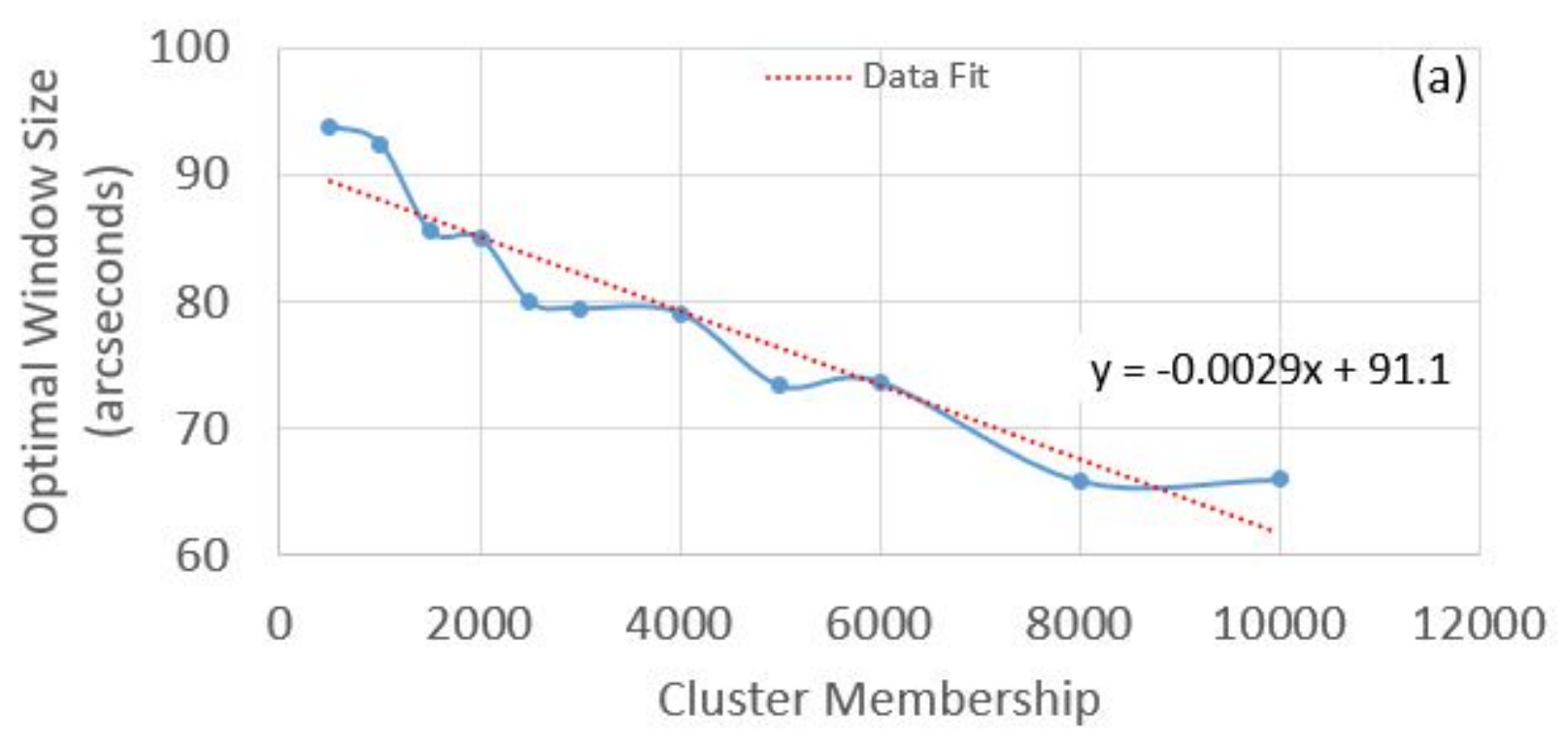}
    \caption{Plot of optimal window size vs cluster membership along with a linear fit for simulated Gaussian clusters without noise.}
    \label{OWS}
\end{figure}

Figure~\ref{OWS} displays the plot of OWS against cluster membership for a given cluster area. As seen from the figure, the OWS varies inversely with the density. In other words, lower the number of stars in the cluster, lower is the density and larger the OWS. The linear fit gives a relation between the two parameters. By varying the cluster radius for a given cluster membership, the same is attained. The larger the cluster size, the lower is the density with a correspondingly larger OWS. For higher density, the window size required is smaller. A larger window size can also be used but this would lead to loss of information at scales smaller than the window size.


Even though we find a roughly linear relationship between window size and density, in the case of real clusters there will be uncertainties as, the distribution of stars in a cluster may not follow a Gaussian. In general, the size of the underlying cluster is unknown in advance, so the algorithm should be run first with a larger window size to efficiently estimate an approximate cluster size. To determine cluster morphology with high accuracy, a window size smaller than the smallest expected cluster is recommended. The latter, however, comes with a higher computational cost. 

\section{Results of Simulated Clusters}
In this section, we describe and discuss the results obtained using the  simulated clusters. Our intention is to quantify and compare the performance of the proposed approach \textit{viz.} Gaussian Parzen Windows, with respect to the other well-known methodologies, such as Star Count, and k-NN.  A cluster without noise, although unrealistic, can be used to distinguish and compare the performance of various algorithms, as the shape and size of cluster is known apriori. Hence, we first discuss the case of the cluster S01, which has the distribution of a Gaussian without background noise.

\subsection{S01: Single Gaussian cluster without background}
In this sub-section, we describe how efficiently each method retrieves the attributes of this cluster. 

\subsubsection{Cluster Size}
\begin{figure*}
    \centering
    \includegraphics[scale=0.58]{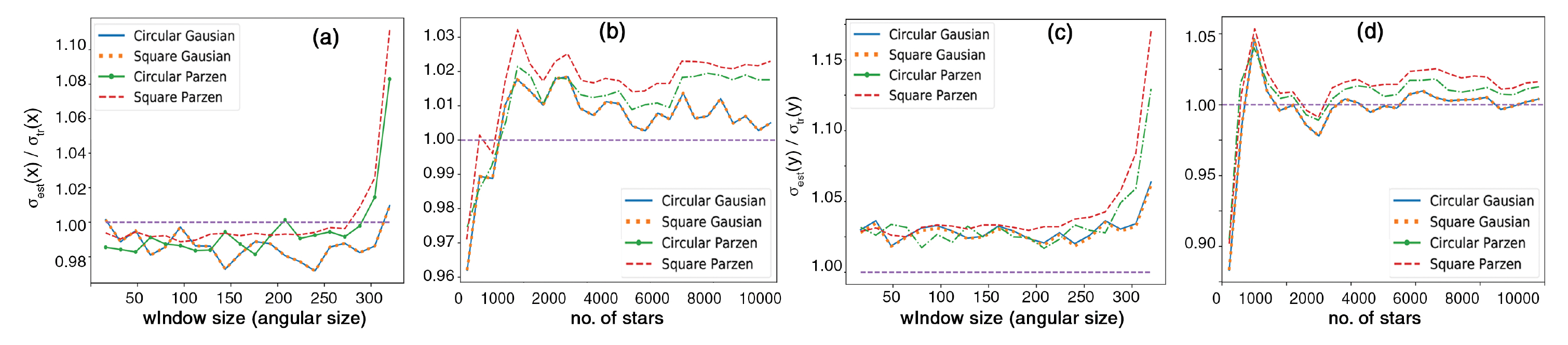}
    \caption{Variation of the ratio of the estimated-to-true size of the 
        single Gaussian Cluster S01: (a) $\sigma_x$ ratio versus size of Parzen window, (b) $\sigma_x$ ratio as a function of size of cluster membership, (c) $\sigma_y$ ratio versus size of Parzen window, (d) $\sigma_y$ ratio as a function of size of cluster membership. }
    \label{Sim_Clus_log_error}
\end{figure*}

For the cluster S01, $\sigma_{tr}$ is already known (being an input parameter), while the $\sigma_{est}$ is extracted using the algorithms. The $\sigma_{est} / \sigma_{tr}$ ratio demonstrates the disparity between the size of the detected cluster with respect to the input cluster. This disparity could be an outcome of under-sampling or limitations of the window(s) employed. For example in the case S01, providing a lower number of datapoints as input to the simulation is likely to result in a distribution which does not represent the Gaussian shape accurately. Hence, the algorithm would not be able to detect the genuine theoretical shape. Similarly, a larger window size would tend to even out the distribution, thus smoothening the shape and modifying the size of the derived Gaussian. As we are utilizing the normalized random number distributions to generate the simulated clusters, altering the seed leads to generation of clusters of the same size but different stellar distributions. This allows for the computation of an average $\sigma_{est} / \sigma_{tr}$ ratio. Figure \ref{Sim_Clus_log_error} displays the variation of this ratio along the two perpendicular directions (denoted by $\sigma_x$ and $\sigma_y$) with respect to clusters of different membership as well as size of the Parzen windows. In Fig. \ref{Sim_Clus_log_error} (a) and (b), the number of stars is kept constant at 1000, while in (c) and (d), the number of windows is kept fixed at 25, that corresponds to a window of size 56.5 angular units.\\

\begin{figure*}
	\centering
	\includegraphics[width=\textwidth]{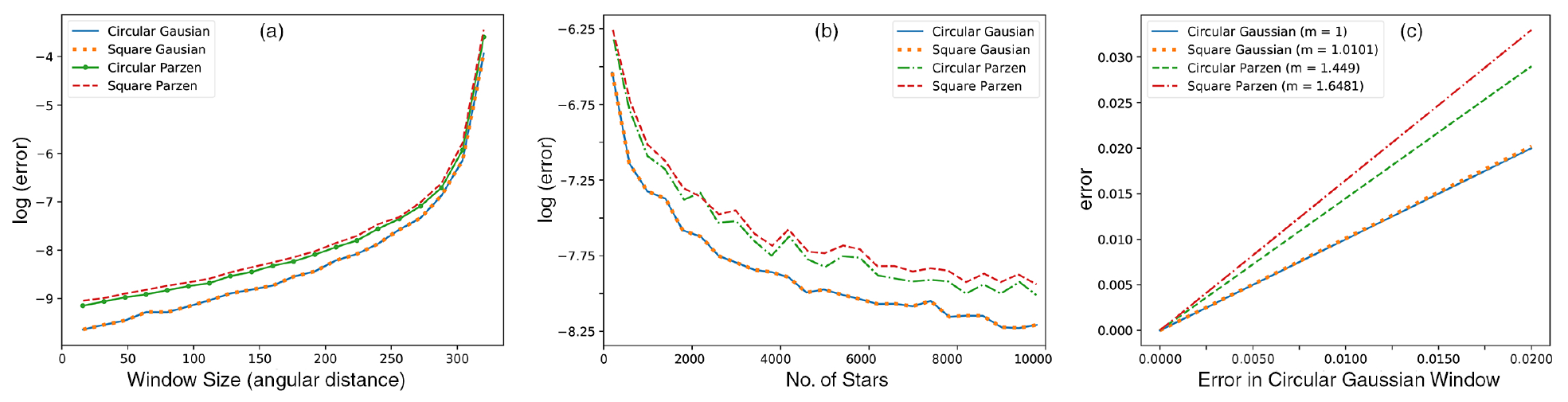}
	\caption{Comparison of Error for S01: (a) log(Error) versus size of Parzen windows, (b) log(Error) as a function of cluster membership, and (c) Window-wise error comparison (see text for details).}
    \label{error_2}
\end{figure*}


By varying the number of Parzen windows (i.e. window size), the square Gaussian had a maximum deviation of 2.7\% and 7.5\% for $\sigma_x$ and $\sigma_y$ whereas the circular Gaussian had a deviation of 2.7\% and 7.3\%, respectively. The maximum deviation was higher for circular and square Parzen windows. The circular Parzen window had a maximum deviation of 8.3\% and 13.3\%, respectively, while the Star Count displayed a maximum of 11.2\% and 16.9\% for the same. Thus, the Gaussian windows approximated the size and shape of the cluster with greater accuracy than the circular and square Parzen windows. The deviation of  the ratio $\sigma_{est} / \sigma_{tr}$, for $\sigma_x$ as well as $\sigma_y$, from unity was larger for smaller number of stars as well as for larger window size, as expected. An increase in the number of stars and windows results in the ratio approaching unity in every case. Although this trend was observed for all window types, for the case of Gaussian Parzen windows, the ratio remains close to unity even for lower number of windows and stars.

\subsubsection{Stellar Distribution in the field}
An absolute element-wise difference, discussed in Eqn. \ref{eq7} measures the error in approximation of the PDF in each window, for a given algorithm. The differences in errors as a function of various parameters (window size and cluster membership) are illustrated in Fig.~\ref{error_2}(a) and (b). These represent the results of simulations with cluster membership as 1000 and window size as 25, respectively. These plots highlight the efficiency of the algorithms in extracting the cluster. In both the cases, we observe that the Gaussian Parzen windows perform better than the other two types. In case of error as a function of number of stars [Fig.~\ref{error_2}(b)], the errors for different window types differ by a constant factor which is highlighted through the slopes of the Window-wise error curves displayed in [Fig.~\ref{error_2}(c)]. The errors obtained for different windows against the error of a circular Gaussian window are shown in Fig.~\ref{error_2}(c) to clearly visualize the comparative errors. Hence, the circular Gaussian curve has a unit slope and serves as a reference line. The square Gaussian versus the circular Gaussian error curve has a slope of a 1.0101. This is in contrast to circular Parzen versus circular Gaussian (slope = 1.4449) and square Parzen versus circular Gaussian (slope = 1.6481). The values of slopes demonstrate that the Gaussian windows yield lower errors than the circular and square Parzen windows. Thus, we infer that the circular Gaussian windows differ the least from the simulated cluster, followed by the square Gaussian and circular Parzen. The Star Count approach performs poorly with respect to the others. \\

\subsection{Other cases of Simulated Clusters}
After considering the simplest case of S01, we proceed to adding background noise to the other mathematically simulated clusters. This allows us to evaluate how efficiently the clusters are detected. In addition, we assess the background level that would enable us to probe the cluster properties discussed earlier. In this section, we present results of the simulated clusters S02 to S06. 
 

\begin{figure}
    \centering
    \includegraphics[width=0.45\textwidth]{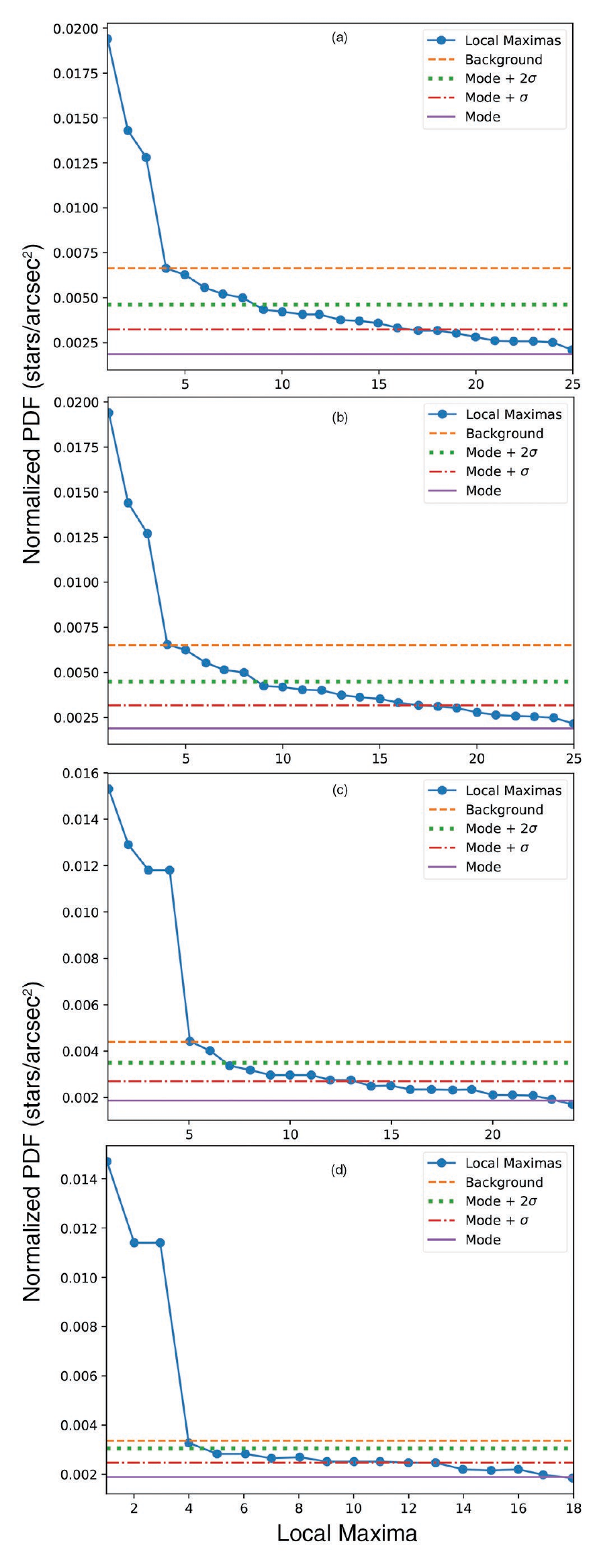}
    \caption{Local Maxima Curves for the simulated cluster S06 using (a) circular Gaussian window, (b) square Gaussian window, (c) circular Parzen window, and (d) square Parzen window. The background limit ($\lambda$) is also marked for each case. }
    \label{locmax_S06}
\end{figure} 


\subsubsection{Background estimation and Cluster identification}

The background level is ascertained based on the method described in Section 4-III. The four algorithms perform well in detecting all the simulated clusters. A comparison of the local maxima curves obtained by the application of the four methods is displayed in Fig.~\ref{locmax_S06} for one of the simulated clusters, S06. This figure illustrates that for a rich to moderately rich cluster, all the windows are able to distinguish the cluster from the noise fluctuations efficiently. However, the local maxima curves obtained using square and circular Gaussian windows are marginally sharper than those obtained using circular Parzen window and Star Count. While this is insignificant for a rich cluster, it becomes crucial for a  cluster with poor membership where the contrast between the cluster and background is not pronounced.

\subsubsection{Cluster Shape and Size}
\begin{figure*}
	\centering
	\includegraphics[width=\textwidth]{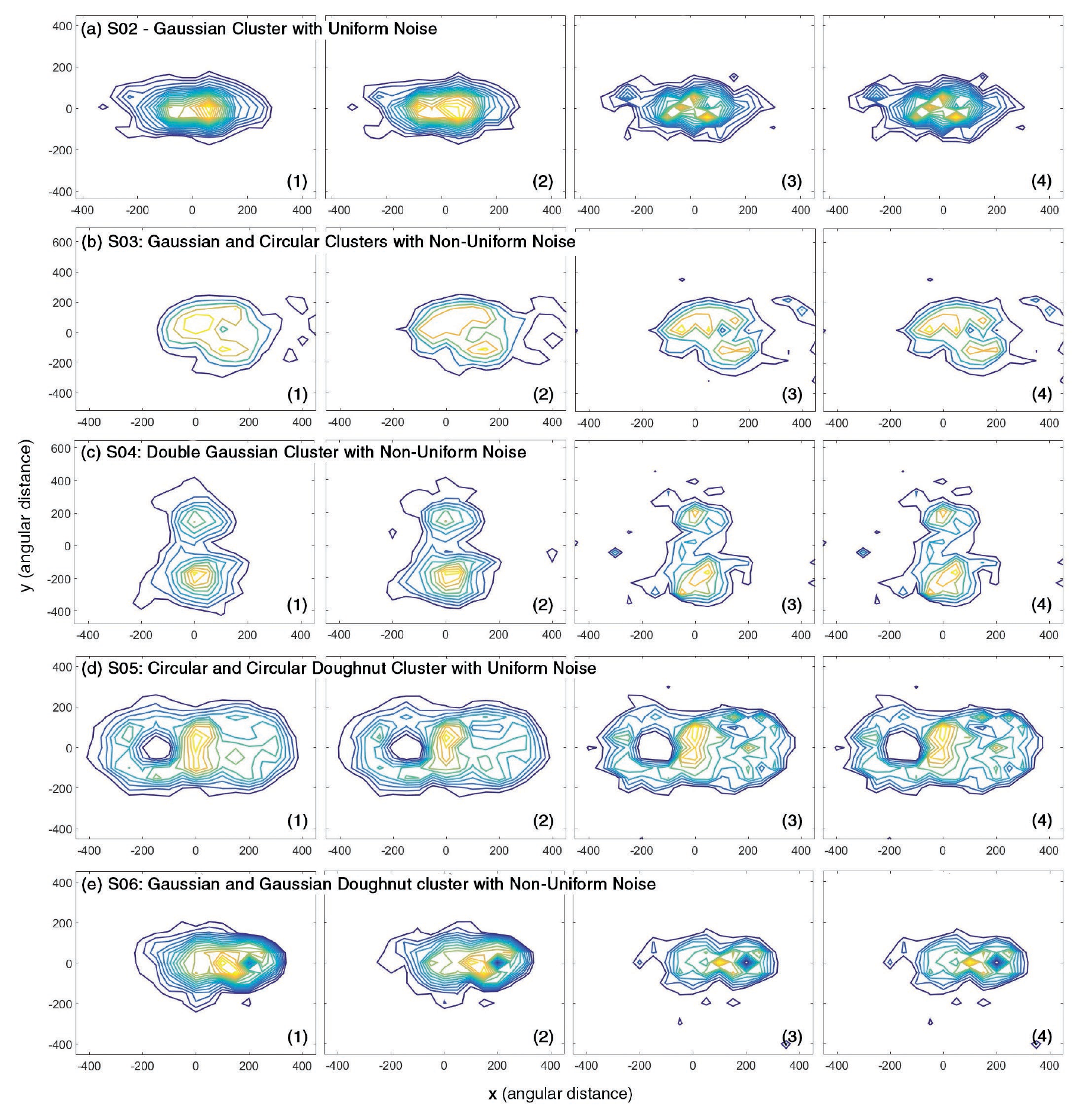}
	\caption{Detection of clusters for the simulated cases using the algorithms (1) square Parzen window, (2) circular Parzen window, (3) square Gaussian window, and (4) circular Gaussian window. The contours begin from the Mode+$2\sigma$ and with an interval of $\sigma$. Each row represents one case of simulated cluster.}
    \label{sim_contour}
\end{figure*}

In order to assess the size of the cluster and comment on its shape, we plot the cluster density through contours. This is shown in Fig.~\ref{sim_contour}. The contours are to be envisaged as loci of points with equal probability density. The peak of the PDF is represented in red and as we move away, the PDF decreases as the colors of contours change gradually to violet. In all the cases of simulated clusters, we see smaller and sharper contours for the Gaussian windows cases whereas we see smoother and larger contours for the other cases, evident from the morphology of the contours. This suggests a loss of information in the latter cases. In particular, this is conspicuous in Fig.~\ref{sim_contour} (d) and (e) for the simulated doughnut clusters S05 and S06, where the void is either missing or is underestimated in size unlike the case of Gaussian windows. It is apparent that the Gaussian windows respond better to fluctuations in density than the square and circular Parzen windows. This emphasizes the superiority of the Gaussian windows in the identification of cluster morphology.

\subsubsection{Number of Stars in the Cluster}
It is possible to compare the true and estimated number of stars in the cluster as the input parameters are known. The differences between the input and derived cluster membership, are listed in Table~\ref{Table:Sim_Clus_error}. These numbers represent the outcome of simulations for a window size of $25''$. We get a maximum error of 17.71\% with the Star Count method for the case of S05, followed by the circular Parzen window approach that gave an over-estimate of 16.73\%. S05 gave the poorest result for these methods among the simulated clusters. Compared to these, the errors for Gaussian windows never exceed 10\%, and in most cases were within 1\% of the true count. We also computed the number of stars obtained by the k-NN approach, for comparison. An optimal value for number of stars in each window was chosen for this purpose which is 20 \citep{schmeja2011001}. The density is estimated using Eqn. \ref{eq5} and the cluster membership evaluated following the procedure outlined for other approaches. By comparing the membership values, we find that the performance of k-NN is relatively inferior or on par with Star Count leading to an under-estimate of 9\%-30\% for different cases, as can be observed from Table \ref{Table:Sim_Clus_error}. In every case, the Gaussian windows behave better than the others. This again highlights the advantage of using the Gaussian windows over the square or circular Parzen windows.


\subsubsection{Stellar Distribution in the field}
The next task is to compare the stellar distributions from the algorithms with the true distribution that is input to the model. Note that this is different from finding the actual number of stars in the cluster. As earlier, we plot the error (i.e. absolute element-wise difference) as a function of number of Parzen windows used. In other words, we change the size of the Parzen window and plot the errors. These are shown in Fig. \ref{error_2}. Again, we obtain differences that vary by constant factors between the corresponding plots of different windows. In these simulated clusters too, it is apparent that the Gaussian windows are able to better approximate the stellar distribution in the area under investigation. Among the other two, circular Parzen windows perform better than the Star Count in all cases. \\

Thus, in every respect, we find that the Gaussian Parzen windows are able to recover the shape of the stellar distribution more accurately than the square and circular Parzen windows, in all the simulated cases. Among the latter, the circular Gaussian performs marginally better. We, therefore, proceed with the hypothesis that the results obtained by using Gaussian windows are more accurate than the simple Parzen windows. 

\begin{figure*}
    \centering
    \includegraphics[scale=0.39]{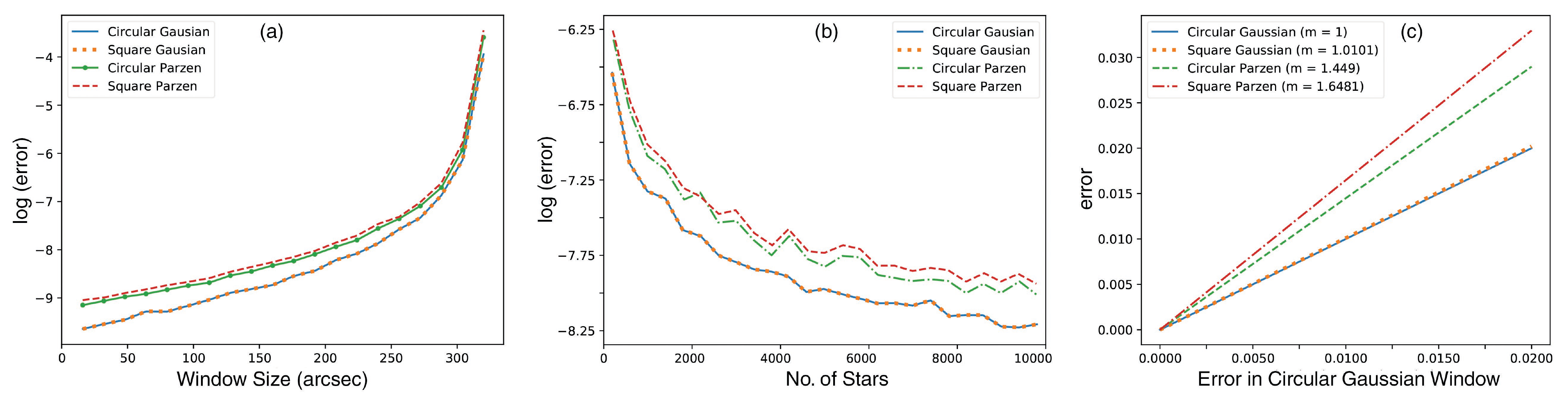}
    \caption{(A) log(Error) vs No. of Parzen  Windows; (B) log(Error) vs No. of Stars (C) Window-wise error comparison for Single Gaussian Cluster S01}
    \label{error_2}
\end{figure*}

\section{Results of Real Clusters}
In this section, we consider regions of the sky where either embedded cluster(s) have been detected or are expected to be present based on other considerations. We use the four algorithms, as previously discussed, to detect the clusters and obtain their properties. Unlike the simulated cases, for a real cluster in general, the parameters are not known in advance. For some of the real clusters considered in this work, although the parameters are known from literature, they are the outcome of the commonly used method, Star Count. As we aspire to compare the performance of Star Count with other methods, we rely on the results of our simulations for comparisons between various methods.

\subsection{Determination of background}

\begin{figure*}
    \centering
    \includegraphics[width=1\textwidth]{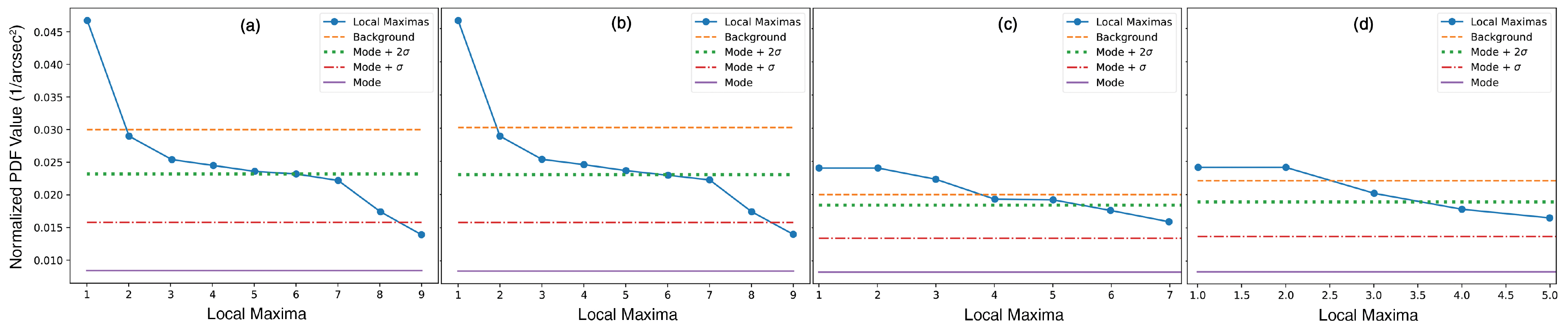}
    \caption{Local Maxima Curve for IRAS 04579+4703 using (a) circular Gaussian window, (b) square Gaussian window, (c) circular Parzen window, and (d) square Parzen window. The background limit ($\lambda$) is also displayed for each case. }
    \label{loc_max_IRAS}
\end{figure*}

In order to segregate the background, we consider the local maxima plots. The local maxima plots for one of the clusters, that is associated with IRAS~04579+4703, are shown in Fig. \ref{loc_max_IRAS}. The figure elucidates that Gaussian windows are better able to distinguish the cluster from the background. In other words, the local maxima curves derived from various algorithms show significant differences in terms of the contrast between the cluster and background. The Gaussian windows exhibit steeper decline in comparison to the square and circular Parzen windows. The latter are averaged versions of the neighbourhood in a sense, leading to an erroneous approximation of the background level. This in turn hinders the estimation of cluster shape and membership. In cases  such as IRAS 01420+6401 (Fig. ~\ref{Real_Clus_contour}(e)) and IRAS 04579+4703 (Fig. ~\ref{Real_Clus_contour}(f)),  the Parzen windows fail to detect the clusters altogether.

\subsection{Cluster Identification and Contours}
The mode+$2\sigma_{bg}$ contour has been used as a threshold for the detection of clusters. 
The detection of clusters by various algorithms is shown through contour plots in Fig.~\ref{Real_Clus_contour}. The clustering is evident in all the regions considered here. The embedded clusters associated with molecular clouds, \textit{viz.} IRAS 01420+6401 and IRAS 04579+4703, were not detected by the application of the Star Count method by \citet{kumar2006001}. Our results show that the square and circular windows could not extract these clusters from the background as the local maxima curves failed to show a significant distinction between the cluster and background levels. These clusters, however, were detected by the algorithms that employ Gaussian windows. It is to be noted that the Star Count method used by us estimates the background threshold from the same field unlike the method employed by \cite{kumar2006001} who used a neighbouring field. The detection of these clusters by Gaussian windows accentuates our claim that Gaussian Parzen windows perform better in cluster detection.

\subsection{Cluster morphology and size}
\begin{figure*}
	\centering
	\includegraphics[width=\textwidth]{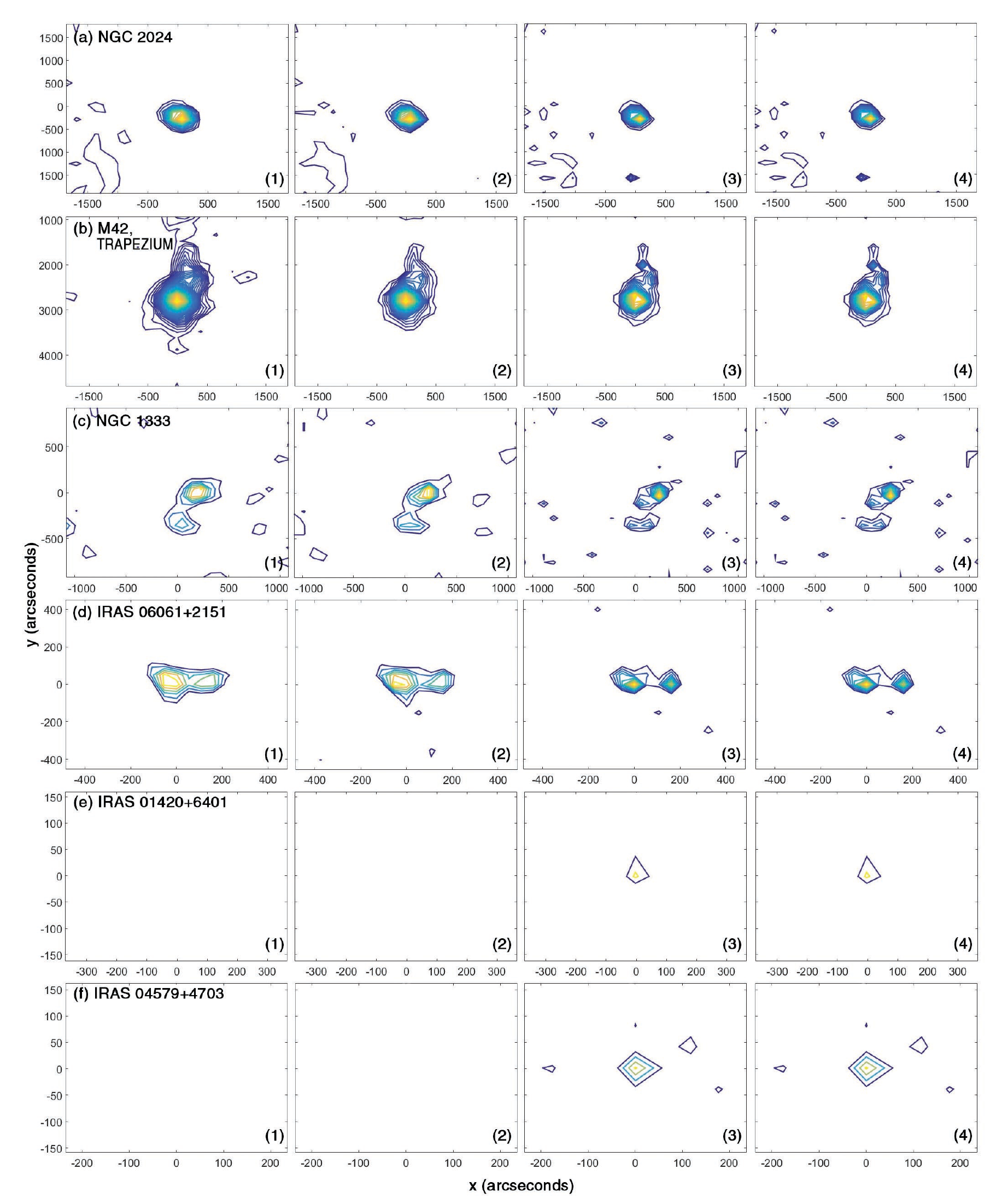}
	\caption{Detection of real clusters for using the algorithms (1) square Parzen window, (2) circular Parzen window, (3) square Gaussian window, and (4) circular Gaussian window. The contours begin from the Mode+$2\sigma$ and with an interval of $sigma$. Each row represents a real cluster.}
    \label{Real_Clus_contour}
\end{figure*}

The cluster morphology is evident from the shape of the contours above the detection threshold. The contours are plotted in Fig.~\ref{Real_Clus_contour}. We clearly perceive that the contours are corrugated for the cases of Gaussian windows as compared to circular and square windows. Based on the results of simulations, this suggests that the smoother contours in the square and circular Parzen windows lead to loss of information in the case of real clusters too. In addition, the Gaussian windows picked up minute details of background in all cases which the square and circular Parzen windows failed to do. And the most compelling result is that, in Fig.~\ref{Real_Clus_contour} (e) and (f), square and circular Parzen windows failed to show any clustering, while Gaussian windows succeeded in revealing the cluster.

We have compared the effective radii of the clusters from various algorithms in Table~\ref{Table:Real_Clus_results2}. In most cases, we discern that the radii revealed by the Gaussian windows are quite similar, unlike the square and circular Parzen windows. The effective radii of these clusters elicited from the literature are also listed in the table. It is important to be cognizant of the fact that a comparison between the values can be tentative at best as the values from literature are derived using the Star Count algorithm by utilising values for background and cluster detection threshold that could be at variance with those considered in this work. In addition, the K-band images used in some of the cited works have sensitivities that differ from that of 2MASS K-band. All these are likely to contribute to uncertainties in the values quoted. 


\subsection{Number of Stars in the Cluster}

The cluster memberships of real clusters are listed in Table~\ref{Table:Real_Clus_results2}. For the rich cluster M42, the estimated figures are very close to each other for all types of windows. The difference increases as the cluster membership decreases, i.e. the signal-to-noise decreases. And for poor clusters  IRAS 01420+6401 and IRAS 04579+4703, no data could be provided by both square and circular Parzen windows. In all the cases, the results of circular and square Gaussian windows are very similar. Similarly, the results of circular Parzen windows and Star Count are also extremely close to each other. Based on the results from simulations, we are compelled to rely more on the cluster membership estimates obtained using the Gaussian windows. The values of cluster membership from literature, using mostly Star Count method, are also listed in Table~\ref{Table:Real_Clus_results2}. While the values are similar, it is to be borne in mind that the images as well as parametric values (eg. background threshold) are likely to be different.

\section{Conclusions}

In conclusion, this paper focuses on a very specific means of detecting star clusters, the Parzen Windows approach. Parzen windows provide a fast and light means to detect star clusters and study its various attributes. Building up on the existing special case of this approach, viz., square Parzen windows or Star Count, this work compares the performance of the square Parzen, circular Parzen, square Gaussian and circular Gaussian windows and demonstrates that the approach of Gaussian windows are superior to the traditional methods of square and circular Parzen windows, in all respects. In addition, we propose a statistical approach to find the background limit and threshold level for cluster detection from the field itself, in contrast to previous studies where this process was based on neighbouring fields. Based on our success with simulations, we apply the Gaussian window algorithms on sample clusters including a couple with low memberships that could not be previously identified using the Star Count method, and successfully detect all. We conclude that the Star Count method with its in-built averaging effect leads to loss of information whereas Gaussian windows retain the small scale features of clusters.

\section*{Acknowledgments}

We are grateful to the referee for suggestions that have improved the quality and presentation of the paper considerably. This work made use of data products from the Two Micron All Sky Survey, which is a joint project of the University of Massachusetts and the Infrared Processing and Analysis Center/California Institute of Technology, funded by the National Aeronautics and Space Administration and the National Science Foundation. This research has made use of the SIMBAD database, operated at CDS, Strasbourg, France.

\begin{landscape}
\begin{table}
  \centering
  \caption{Results of Real Clusters}
    \begin{tabular}{|c|c|ccc|cc|cc|cc|cc|}
    \toprule
    \textbf{S. No.} & \textbf{Cluster Name} & \textbf{Cluster Center} & \textbf{Size of Region} & \textbf{WS} & \multicolumn{2}{c|}{\textbf{SP}} & \multicolumn{2}{c|}{\textbf{CP}} & \multicolumn{2}{c|}{\textbf{SG}} & \multicolumn{2}{c|}{\textbf{CG}} \\
\cmidrule{6-13}          &       & \textbf{($\Delta\alpha$,$\Delta\delta$)} &	 &       &  \textbf{Mode} & \textbf{DT} & \textbf{Mode}  & \textbf{DT} & \textbf{Mode} & \textbf{DT} & \textbf{Mode} & \textbf{DT} \\
          &       &       &       &       & \textbf{(s/a$^2$)} & \textbf{(s/a$^2$)} & \textbf{(s/a$^2$)} & \textbf{(s/a$^2$)} & \textbf{(s/a$^2$)} & \textbf{(s/a$^2$)} & \textbf{(s/a$^2$)} & \textbf{(s/a$^2$)} \\
    \midrule
    1     & NGC 2024 & (85.433, -1.823) & 2000$''$ x 2000$''$ & 153.84$''$ & 0.0012 & 0.0025 & 0.0012 & 0.0026 & 0.0012 & 0.0028 & 0.0012 & 0.0028 \\
    2     & NGC 1333 & (52.3099, 31.4003) & 1000$''$ x 1000$''$ & 166.92$''$ & 0.0009 & 0.0029 & 0.0009 & 0.0034 & 0.0009 & 0.0039 & 0.0009 & 0.0039 \\
    3     & M42 Trapezium & (83.8376, -4.609) & 4000$''$ x 4000$''$ & 258.06$''$ & 0.0008 & 0.0013 & 0.0008 & 0.0017 & 0.0008 & 0.0023 & 0.0008 & 0.0023 \\
    4     & IRAS 06061+2151 & (92.2975, 21.8546) & 500$''$ x 500$''$ & 98.09$''$ & 0.0029 & 0.0046 & 0.0029 & 0.0047 & 0.0029 & 0.0053 & 0.0029 & 0.0053 \\
    5     & IRAS 01420+6401 & (26.4136, 64.2674) & 200$''$ x 200$''$ & 72.72$''$ & -     & -     & -     & -     & 0.015 & 0.0254 & 0.015 & 0.0254 \\
    6     & IRAS 04579+4703 & (75.4396, 47.1251) & 201$''$ x 200$''$ & 72.72$''$ & -     & -     & -     & -     & 0.0084 & 0.0224 & 0.0084 & 0.0225 \\
    \bottomrule
    \end{tabular}
  \label{Table:Real_Clus_results}
  \begin{tablenotes}
      \item \textit{s/a$^2$} = stars/arcsec$^2$
      \item \textit{SP} = Square Parzen
      \item \textit{CP} = Circular Parzen
      \item \textit{SG} = Square Gaussian
      \item \textit{CG} = Circular Gaussian
      \item \textit{WS} = Window Size
      \item \textit{DT} = Detection Threshold
    \end{tablenotes}
\end{table}


\begin{table}
  \centering
  \caption{Comparison of cluster parameters with Literature}
    \begin{tabular}{|c|c|ccccc|ccccc|}
    \toprule
    \textbf{S. No.} & \textbf{Cluster Name} & \multicolumn{5}{c|}{\textbf{Cluster Membership}} & \multicolumn{5}{c|}{\textbf{Effective Radius}} \\
\cmidrule{3-12}       &   & \textbf{SP} & \textbf{CP} & \textbf{SG} & \textbf{CG} & \textbf{Lit} & \textbf{SP} & \textbf{CP} & \textbf{SG} & \textbf{CG} & \textbf{Lit} \\
\cmidrule{1-12}    1 &  NGC 2024  & 893   & 855   & 687   & 687   &  309$^1$ & 585.16$''$ & 585.16$''$ & 518.7$''$ & 518.7$''$ & $454''^1$ \\
    2 &  NGC 1333  & 176   & 147   & 132   & 132   & 143$^2$ & 341.74$''$ & 288.83$''$ & 296.96$''$ & 288.83$''$ & $318''^2$ \\
    3 &  M42 Trapezium  & 3099  & 2369  & 1988  & 1988  & 1740$^3$ & 836.4$''$ & 603.3$''$ & 488.7$''$ & 488.7 & $1742''^3$ \\
    4 & IRAS 06061+2151   & 145   & 139   & 122   & 122   & 49$^{4}$ & 116.97$''$ & 124.07$''$ & 109.41$''$ & 109.41$''$ & $145''^{4}$ \\
    5 &  IRAS 01420+6401  & -     & -     & 4     & 4     &       & -     & -     & 34.25$''$ & 34.25$''$ & - \\
    6 &  IRAS 04579+4703  & -     & -     & 19    & 18    &       & -     & -     & 66.93$''$ & 66.93$''$ & - \\
    \bottomrule
    \end{tabular}
  \label{Table:Real_Clus_results2}
    \begin{tablenotes}
      \item \textit{Lit} = Literature
      \item \textbf{ \textit{Rest of the abbreviations are same as in Table~\ref{Table:Real_Clus_results}}}
    \end{tablenotes}
\begin{flushleft}
$^1$ \citet{lada1991002} \\
$^2$ \citet{ladalada1996001} \\
$^3$ \citet{ladalada2003001} \\
$^4$ \citet{kumar2003001} \\
\end{flushleft}
\end{table}
\end{landscape}

\bibliographystyle{mnras}
\bibliography{mybibfile}

\bsp

\label{lastpage}

\end{document}